\documentclass[]{spie}  

\usepackage{amsmath,amsfonts,amssymb}
\usepackage{graphicx}
\usepackage{subfig}
\usepackage[colorlinks=true, allcolors=blue]{hyperref}

\title{Comparing linear structure-based and data-driven latent spatial representations for sequence prediction}

\author[a,b]{Myriam Bontonou}
\author[a,b]{Carlos Lassance}
\author[a,b]{Vincent Gripon}
\author[a]{Nicolas Farrugia}
\affil[a]{IMT Atlantique and Lab-STICC, France}
\affil[b]{Universit\'e de Montr\'eal and MILA, Canada}

\authorinfo{Further author information:\\ M.B.: myriam.bontonou@imt-atlantique.fr,    C.L.: carlos.rosarkoslassance@imt-atlantique.fr,\\ V.G.: vincent.gripon@imt-atlantique.fr,    N.F.: nicolas.farrugia@imt-atlantique.fr.}

\begin{document} 
\maketitle

\begin{abstract}
Predicting the future of Graph-supported Time Series (GTS) is a key challenge in many domains, such as climate monitoring, finance or neuroimaging. Yet it is a highly difficult problem as it requires to account jointly for time and graph (spatial) dependencies.
To simplify this process, it is common to use a two-step procedure in which spatial and time dependencies are dealt with separately.
In this paper, we are interested in comparing various linear spatial representations, namely structure-based ones and data-driven ones, in terms of how they help predict the future of GTS. 
To that end, we perform experiments with various datasets including spontaneous brain activity and raw videos.
\end{abstract}


\section{INTRODUCTION}
\label{sec:intro}
Spatially structured temporal sequences are everywhere: in video prediction, in climate monitoring, in functional Magnetic Resonance imaging (fMRI) and so on. Although much progress has been made (c.f. Section~\ref{sec:rw}), predicting their evolution over time remains a highly difficult challenge.

Such sequences are naturally described as signals over a graph (or a series of graphs), evolving over time. For instance, in the case of fMRI signals, the nodes of the graph are groups of neurons called Regions Of Interest (ROI), the spatial structure is described by the edges of the graph, which represent either the physical connections or statistical similarities between signals across ROIs. The fMRI signal is an indirect measure of brain activity through the blood-oxygen level dependent (BOLD) signal~\cite{ogawa1990brain}. Averaging such signals in ROIs results in a scalar for each node of the graph at each time point.

The prediction of the evolution of such sequences in the raw spatial domain can prove to be very challenging, especially if the spatial domain is subject to important noise.
Instead, we would prefer to predict the future of these sequences in a simplified latent space representation, in which the content is described using sparse, robust, low-dimensional representations.

Finding such representations is a common problem in signal processing and data science. The most popular methods are principal components analysis and independent component analysis. In signal processing, recent research efforts attempt at exploiting graphs to derive such low-dimensional representations, by defining the Graph Fourier Transform (GFT), or the spectral graph wavelet transform~\cite{pilavci2019spectral}. In the context of optimization, popular methods include dictionary learning~\cite{mairal2009online} and non-negative matrix factorization~\cite{lee1999learning}. In machine learning,  auto-encoders and their variants are popular frameworks based on neural networks to learn latent representations~\cite{bengio2013representation}. 
Because it simply consists in changing the basis in which data is represented, GFT comes with a rich set of guarantees, theorems and intuitions. And as such, it is a strong candidate in a rising field of research~\cite{shuman2012emerging}.
However, it is known that GFT can fail at fully leveraging the underlying structure, even in the case of very regular domains~\cite{lassance2018matching}.

The question we ask in this paper is: what are the linear transforms that are the best fitted to predict the evolution of GTS?

We consider three approaches. First, we introduce a fully structured-based approach, where the linear transform is defined using the lower part of the spectrum of the Graph Fourier basis. Second, we consider a mixed strategy, where both the structure and data are considered while looking for the best basis. To this end, we use a semi-geometric graph which support is the same as for the first case, but where the weights of the edges account for the covariance of the signals measured at different nodes. Finally, we consider a learned scheme, where the structure is disregarded and only data helps in finding the good representation. 

The question we ask is twofold:
\begin{enumerate}
    \item Which linear transforms give the best latent representation for spatial compression of the signals?
    \item Which linear transforms give the most fitted representations for sequence (time) prediction?
\end{enumerate}

As we consider data-driven approaches in our study, we mainly support our claims through experimental validation. We thus consider three datasets made of images and fMRI signals.

The outline of the article is as follows. In Section~\ref{sec:rw}, we introduce related works. In Section~\ref{sec:method}, we explain our methodology to obtain latent spatial representations and perform sequence (time) prediction. In Section~\ref{sec:result}, we empirically compare different approaches. Finally, Section~\ref{sec:discuss} is a discussion and Section~\ref{sec:ccl} a conclusion.

\section{RELATED WORK}
\label{sec:rw}
Many studies so far have dealt with temporal prediction of spatially structured data. For instance, in deep learning, variants of Recurrent Neural Networks (RNNs), such as Long-Short Time Memory (LSTM)~\cite{hochreiter1997long}, are the basis of temporal prediction. These networks build an internal memory about previous data. Thus, they are able to remember previous information to inform later steps. Data is input frame after frame in the networks. At each step, they output an estimate of the future frame. However, the networks do not handle directly data structure. In past years, some works addressed the problem. For instance, in video prediction, data is images. Replacing the RNNs internal linear operations by convolutions allows them to handle the image structure. It has been shown that it improves frame prediction~\cite{xingjian2015convolutional, wang2017predrnn, wang2018eidetic}. More recently, prediction on data residing on the nodes of graphs have received much attention. A basic idea is to replace the RNNs internal linear operations by graph convolutions, but the best results are achieved using more complex architectures~\cite{li2017diffusion, zhang2018gaan}.

In Deep Learning, most methods act as black boxes in which parameters are tuned using gradient descent, without the ability to understand the purpose of each of them. Finding the good hyperparameters can be very demanding, as the combinatorial search space rapidly explodes. Therefore, it is hard to obtain guarantees about robustness of the method, or even reproducibility on other datasets.

When it comes to graphs, a natural approach to define a latent representation is a structure-based approach. It consists in defining the notion of frequency on a graph by analogy with the Fourier Transform (FT)~\cite{shuman2012emerging}. The so-called Graph Fourier Transform (GFT) represents the signal in another basis. This basis is built according to the graph structure without looking at the data at all. The eigenvectors of the Laplacian matrix of the graph are used as a new basis. These eigenvectors are ordered by increasing eigenvalues. Small eigenvalues correspond intuitively to low frequencies, and large eigenvalues to high frequencies. Consequently, to reduce the data dimension, only some of the first eigenvectors are kept. Data is back projected using only those first eigenvectors, and a low-dimensional representation of the signal is obtained. However, the analogy with the FT is not complete, as for instance, the GFT using a 2-dimensional grid does not match the 2D classical FT~\cite{lassance2018matching}. 

Finding the best graph structure for a given problem is a question that recently sparked a lot of interest in the literature~\cite{pasdeloup2017characterization,dong2016learning,marques2017stationary}. In many cases, a good solution consists in using the covariance matrix (or its inverse) as the adjacency matrix of the graph. Following this lead, in this paper, we make use of a semi-geometric graph. It combines the geometry of images captured through a regular grid graph with the covariance of the pixels measured on different frames. As such, the support of the graph remains the same as that of the grid graph, but the weights are replaced by the covariance between corresponding pixels. Therefore, this graph takes into account both the data structure and the data distribution.

Another way to represent data in a low-dimensional space is to use an auto-encoder. An auto-encoder is a deep neural network that is trained to reproduce its input as output, going through an intermediate latent representation. As such, the first part of the auto-encoder compresses data in a low-dimensional space. The second part decompresses the latent representation to evaluate how well it represents the raw data. A variety of complex non-linear auto-encoders have been proposed~\cite{bengio2013representation, chen2016variational}, but for a fair comparison, in this article we only use a simple variant that mimics the principles of GFT.

In this work, much of the interest in the latent representation is motivated by its use for sequence time prediction. There are, again, several recent works in this area, based on diverse methods (dictionary learning~\cite{la2018multivariate}, source separation~\cite{hjelm2018spatio}). Again, we restrict the study to a simple question: what is the best linear, structure-based or data-driven, representation for predicting future frames using a LSTM?

\section{METHODOLOGY}
\label{sec:method}
We consider spatially structured temporal sequences of data. These data are intra-connected. Let us take our two examples, images and fMRI signals. Images are spatially structured in the sense that each of their pixel correspond to a position inside the image. These positions can be neighbors or to the contrary far away. Similarly, fMRI signals are spatially structured, as these signals are associated with a group of neurons (called regions of interest (ROIs)). Each ROI is physically connected to other ROIs.

In both cases a simple way to represent data structure is to use a graph. Each variable (pixels, ROIs) is associated with a node of the graph. Then, connections are drawn between these variables and represented by edges in the graph. The value of each variable is seen as a signal (i.e. a vector) where each dimension corresponds to a node in the graph.

To perform (time) sequence prediction, a simple way is to train a Fully-Connected Long Short-Term Memory (FC-LSTM) neural network on it. But it can prove to be challenging (memory loss) as it does not take into account data structure. Here, we are interested in modifying the input of the FC-LSTM to make it reflect data structure. We begin with the simpler question of finding a linear spatial latent representation. Then, we investigate whether this latent representation helps the FC-LSTM provide more accurate predictions.

\subsection{Comparing linear latent representations of spatially structured data}
In this section, we forget the temporal aspect of the sequences. We only evaluate the ability of linear methods to effectively compress, and then reconstruct, the raw data.

Let $G = \langle V, \mathbf{W}\rangle$ be a weighted graph with $n$ nodes. $V = \{1,\dots,n\}$ is the finite set of $n$ nodes and $\mathbf{W}$ is the weighted adjacency matrix. $\mathbf{W}_{ij}$ is the weight of the edge between nodes $i$ and $j$, or 0 if no such edge exists. $\mathbf{x}\in\mathbb{R}^n$ is a signal on this graph, so that its $i$-th component corresponds to node $i$ in $G$.

A structure-based approach used on data reposing on graphs is the \emph{Graph Fourier Transform} (GFT)~\cite{shuman2012emerging}. This approach builds an analogy with the Fourier Transform. Let us consider the (combinatorial) Laplacian matrix $\mathbf{L} = \mathbf{D} - \mathbf{W}$, where $\mathbf{D}$ is the diagonal degree matrix: $\mathbf{D}_{ii} = \sum_{j=1}^{n}{\mathbf{W}_{ij}}$. Provided that $\mathbf{W}$ is real symmetric, it admits a set of orthonormal eigenvectors $\left \{\mathbf{u}_l \right \}_{l=1}^{n}\in\mathbb{R}^n$. We denote $\mathbf{U}=[\mathbf{u}_1,...,\mathbf{u}_{n}]\in\mathbb{R}^{n\times n}$ and $\mathbf{U}^T$ the transpose of $\mathbf{U}$.
By analogy with the FT, the GFT of the signal $\mathbf{x}$ is defined as: 
$\hat{\mathbf{x}} = \mathbf{U^T}\mathbf{x}$
and the inverse GFT is defined as: $\mathbf{x} = {\mathbf{U}}\;\hat{\mathbf{x}}$. 
From that, we want to compute a latent representation with $m$ latent variables. By taking only a subset of eigenvectors, the ones associated with the $m$ lowest eigenvalues, we build a latent representation of the data. The complexity of this method is in general cubic with the number of nodes in the graph, since it requires diagonalizing a matrix.

In the case of fMRI signals, we use a graph computed from the correlation matrix of the measurements between the ROIs over time, thresholded to keep only the edges with the 5~\% largest weights. In the case of images, we consider two kind of graphs. the first one is a grid graph, completely agnostic of the data. In this graph, each pixel is connected to its neighbors pixels (left, right, up, down). The edge weight is either 0 (no connection) or one (connection). The other graph is a semi-geometric graph. This graph has the same support as a grid but the edge weight is now the covariance between the pixels values over time.

To be able to use GFT requires to have access to a graph describing the spatial domain of the data. Such graphs are not always accessible. An alternative is to use a data-driven approach. In this work we choose to use auto-encoders. For a fair comparison, we define a very basic version of an auto-encoder. Generally speaking, an auto-encoder is split into two parts: an encoder and a decoder. In this study, the encoder is a linear transformation. Formally, given the signal $\mathbf{x}\in\mathbb{R}^n$ and a weight matrix $\mathbf{A}\in\mathbb{R}^{n \times m}$, the encoder is defined as $\hat{\mathbf{x}} = \mathbf{A}^T\mathbf{x}$. The decoder uses the transposed matrix $\mathbf{A}$ to reconstruct the signal $\tilde{\mathbf{x}} = {\mathbf{A}}\;\hat{\mathbf{x}}$. Thus, its mathematical formulation is very similar to that of the GFT, with the noticeable difference that it does not require priors about the spatial structure of data.

The Mean Square Error (MSE) between raw images and reconstructed images is used to train the AE. It is also used to evaluate the accuracy of all methods. 

\subsection{Does the latent representation help in having better predictions?}
Once latent representations have been obtained, using either the GFT or an auto-encoder (AE), the next step is to use them to predict the future of a sequence. In this work, we use a FC-LSTM to predict the evolution of sequences of data. Spatially compressed sequences are input into the FC-LSTM frame by frame. For each element $\mathbf{x}_t \in\mathbb{R}^{m}$, two functions are updated: the cell state $c_t$ and the hidden state $h_t$. The cell state acts as the memory of FC-LSTM. The hidden state corresponds to the prediction of the next element. More precisely, at time $t$, the input, forget, cell and output gates are computed, respectively $\mathbf{i}_t$, $\mathbf{f}_t$, $\mathbf{c}_t$ and $\mathbf{o}_t$. All of them are vectors in $\mathbb{R}^{m}$.
$\mathbf{W}_{..}$ and $\mathbf{b}_{..}$ are trainable parameters. $\sigma$ is the sigmoid function. $*$ is the Hadamard product.

\begin{equation}
\label{eq:lstm}
  \begin{split}
    \mathbf{i}_t = \sigma(\mathbf{W}_{ii}\mathbf{x}_t + \mathbf{b}_{ii} + \mathbf{W}_{hi}\mathbf{h}_{t-1} + \mathbf{b}_{hi}),
    \\ \mathbf{f}_t = \sigma(\mathbf{W}_{if}\mathbf{x}_t + \mathbf{b}_{if} + \mathbf{W}_{hf}\mathbf{h}_{t-1} + \mathbf{b}_{hf}),
    \\ \mathbf{g}_t = \tanh(\mathbf{W}_{ig}\mathbf{x}_t + \mathbf{b}_{ig} + \mathbf{W}_{hg}\mathbf{h}_{t-1} + \mathbf{b}_{hg}),
    \\ \mathbf{o}_t = \sigma(\mathbf{W}_{io}\mathbf{x}_t + \mathbf{b}_{io} + \mathbf{W}_{ho}\mathbf{h}_{t-1} + \mathbf{b}_{ho}),
    \\ \mathbf{c}_t = \mathbf{f}_t * \mathbf{c}_{t-1} + \mathbf{i}_t * \mathbf{g}_t,
    \\ \mathbf{h}_t = \mathbf{o}_t * tanh(\mathbf{c}_t).
  \end{split}
\end{equation}

Sequences are split into two parts: the first is used to warm up the FC-LSTM and the second to evaluate its prediction ability. Indeed, at the beginning, the cell state and the hidden state are initialized at zero. Then, for each frame of the first part, the frame is input into the FC-LSTM which outputs a prediction. Finally, for each frame of the second part, the previous prediction is input into the FC-LSTM which again outputs a prediction.

The Mean Square Error (MSE) between frames and predictions is used to train the FC-LSTM. However, for the evaluation, only the MSE between frames and predictions of the second part is displayed in the results.

\subsection{Datasets}
In this study, we use three datasets, namely MovingSTL, MovingMNIST and a dataset of resting state fMRI. 

MovingSTL is built from STL10~\cite{coates2011analysis}. STl10 images are $ 96\times 96 $ pixels. We transform these images into grey images, normalized between [-1, 1]. We use 5000 images. 3500 for training. 1500 for testing. Each image is used to define a sequence. A square, $ 45\times 45 $ pixels, is randomly cropped in the image. This square randomly moves by one pixel 20 times in a random direction. Thus, we define 5000 sequences of 20 images of $ 45\times 45 $ pixels.

For MovingMNIST~\cite{szeto2018dataset}, we pick 5000 sequences of 20 images. 3500 for training. 1500 for testing. Each image is $ 64\times 64 $ pixels. A sequence displays a moving white number on a black background.

The fMRI dataset is a subset of 62 subjects from the MPILMBB resting-state dataset~\cite{mendes2019functional}. Participants were measured four times at rest with eyes-open, fixating a cross during
approximately 15 minutes (TR = 1400 ms; voxel size = 2.3 mm isotropic). Here, we only use the first measurement for each subject. 
Preprocessing and denoising are described with more details here \url{https://neuroanatomyandconnectivity.github.io/opendata/}. The denoised data was z-scored and normalized to MNI space and parcellated on 523 non-overlapping ROIs from the finest scale of BASC atlas (444 networks)~\cite{bellec_multi-level_2010}. Two subjects were rejected because of corrupted data for one subject (the time series was shorter than expected), and because of excessive head motion for the other subject. 60 subjects remain, and we use 48 subjects for training and the 12 remaining subjects for testing.

On MovingSTL and MovingMNIST, the auto-encoder is trained for 400 epochs using 100 examples per mini-batch, with Adam, using a learning rate of 1e-5 with a weight decay factor that starts at 1e-5 and is divided by 10 at epochs 4 and 120. On the fMRI signals, the auto-encoder is trained for 400 epochs using 6 examples per mini-batch, with Adam, using a learning rate that starts at 1e-5, and is divided by 2 at epoch 200 with a weight decay factor of 1e-5. Only MovingSTL and the fMRI signals are used for evaluating whether the latent representation helps in having better predictions. On both, the FC-LSTM is trained for 600 epochs using 6 examples per mini-batch, with Adam, using a learning rate that starts at 0.001 and is divided by 2 at epochs 200 and 400, with no weight decay.

\section{RESULTS}
\label{sec:result}
\subsection{Linear latent representations of spatially structured data}
\begin{figure}[b]
  \begin{center}
    \subfloat[MovingSTL test set]{{\includegraphics[width=7.5cm]{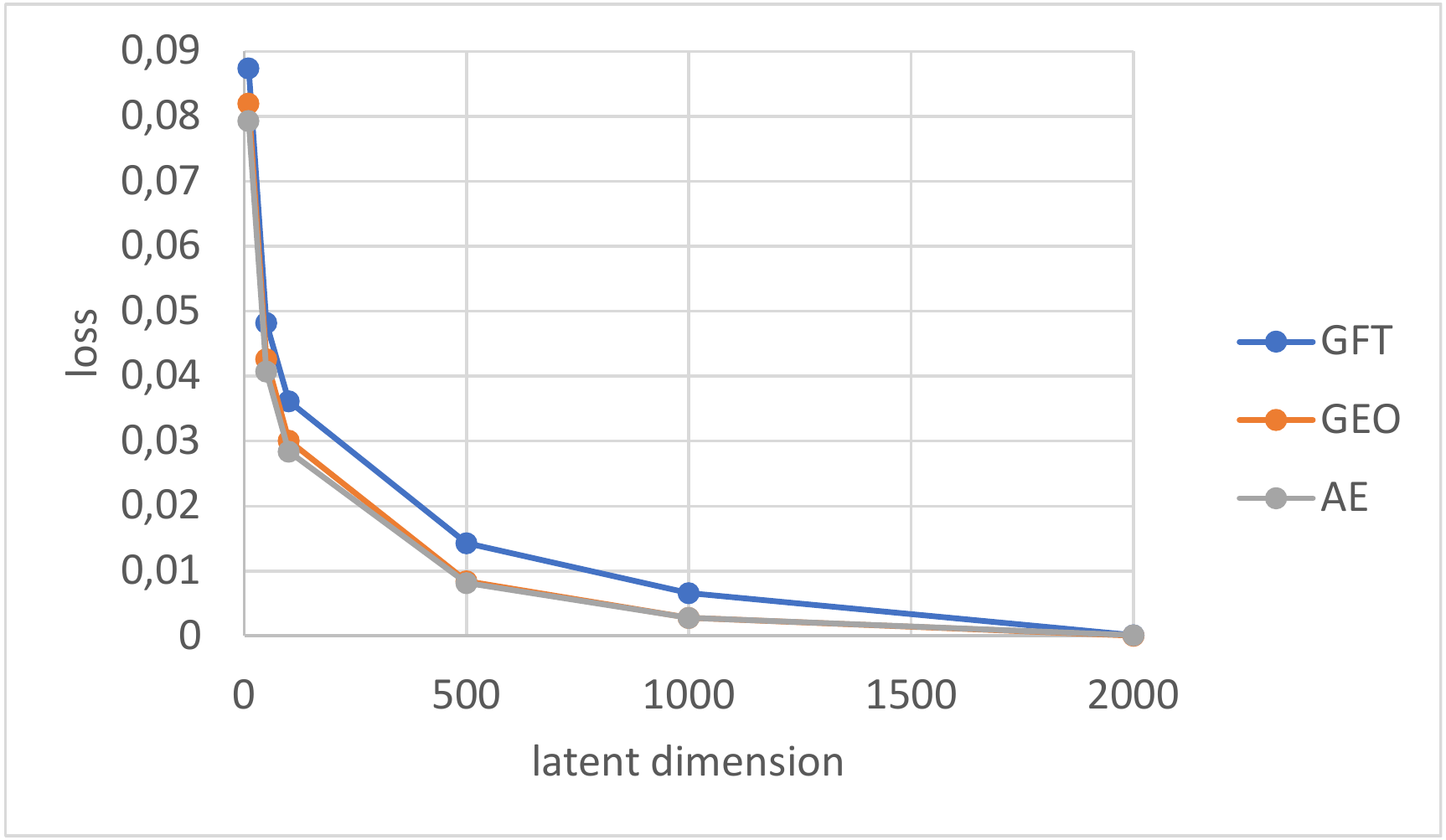}}}
    \qquad
    \subfloat[MovingMNIST test set]{{\includegraphics[width=7.5cm]{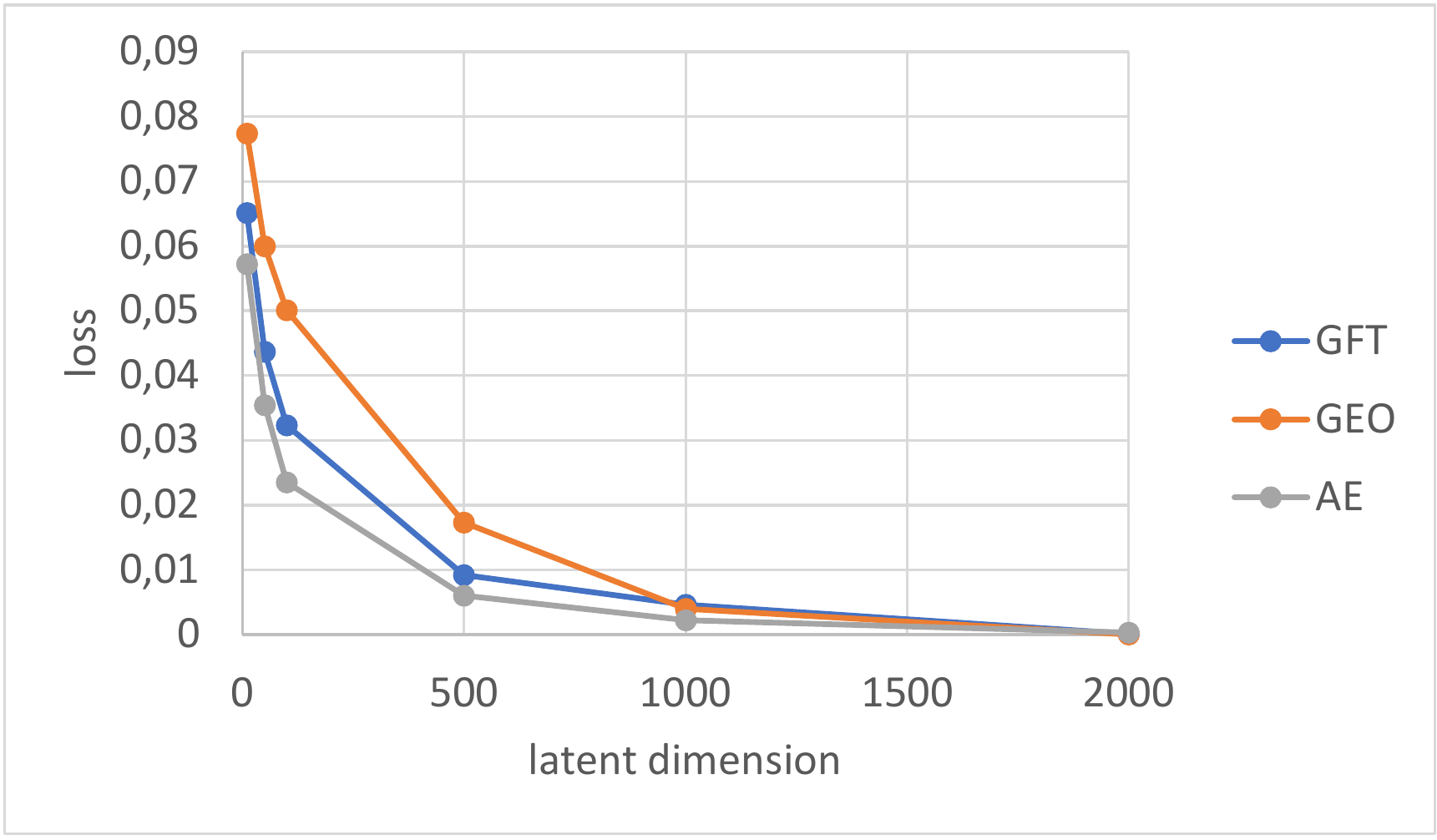}}}
  \caption{\label{Fig:latent_STL} MSE loss as a function of the number of latent dimensions $m$}
  \end{center}
\end{figure}

In the case of images, we evaluate the ability of our three linear transforms to reconstruct raw images from their compressed representations. The three transforms are: the Graph Fourier Transform applied on a grid graph (GFT), the Graph Fourier Transform applied on a semi-geometric graph (GEO) and a linear auto-encoder (AE). We compute the MSE between raw images, from MovingSTL and MovingMNIST, and their reconstruction in function of the number of latent representations. See Fig.~\ref{Fig:latent_STL}.
For MovingSTL, the figure shows that, whatever the latent dimension, the AE reconstructs better the raw data than GFT. The results of AE and GEO are very similar. For MovingMNIST, the AE is also better but the GEO results are worse. 
In the case of fMRI data, we evaluate the ability of the Graph Fourier Transform applied on a correlation graph (GFT) and a linear auto-encoder (AE). We compute the MSE between raw data and their reconstruction in function of the number of latent dimensions for 523 regions of interest (ROI) and for 171 ROI. See Table~\ref{table:latent_fMRI}. Just to notice it, the data for 171 ROI has been obtained after a hierarchical clustering of the 523 ROI. Again, the AE obtains a better MSE, whatever the latent dimension.

\begin{table}[b]
\begin{center}
\subfloat[171 ROI]{{

\begin{tabular}{|c|c|c|}
\hline
                      & \multicolumn{2}{c|}{Loss} \\
\hline
Latent dimension      & 86   & 43 \\
\hline
GFT                   & 0.126   &  0.246        \\
AE                    & 0.065   &  0.134         \\
\hline
\end{tabular} }}
\qquad
\subfloat[523 ROI]{{
\begin{tabular}{|c|c|c|}
\hline
                      & \multicolumn{2}{c|}{Loss} \\
\hline
Latent dimension      & 262   & 131 \\
\hline
GFT                   & 0.174   &  0.317        \\
AE                    & 0.102   &  0.189         \\
\hline 
\end{tabular} }}
\end{center}
\caption{MSE loss on the fMRI test set as a function of the number of ROI and the number of latent dimensions.}
\label{table:latent_fMRI}
\end{table}

\subsection{Predicting the evolution of the sequences with latent representations}
In this section, we evaluate the ability of a FC-LSTM to predict the dynamics of GTS from compressed representations.
In the case of MovingSTL, for a given number of latent dimensions, the prediction MSE for the three linear transforms is similar. See Table ~\ref{table:lstm-STL}. As the reconstruction errors of the three methods are not the same, it raises questions. We discuss this result in the next section. We also notice that their predictions are better than the ones obtained without latent representations. Appendix~\ref{sec:pred} contains two examples of predictions.
In the case of the fMRI measurements, unfortunately, we were not able to obtain accurate predictions. Whatever the method, the prediction converges quickly towards the average signal value. It is not surprising as our FC-LSTM is only able to predict the structure of a few frames before it starts to converge towards an average value as well.

\begin{table}[t]
\begin{center}
\subfloat[Latent dimension: 1000]{{
\begin{tabular}{|c|c|c|}
\hline
         & Prediction MSE & Reconstruction MSE \\
\hline
GFT                   & 0.094   & 0.007    \\
GEO                   & 0.093   & 0.003   \\
AE                    & 0.096   & 0.003   \\
\hline
\end{tabular}}}
\qquad
\subfloat[Latent dimension: 500]{{
\begin{tabular}{|c|c|c|}
\hline
       & Prediction MSE & Reconstruction MSE  \\
\hline
GFT                   & 0.093   & 0.014    \\
GEO                   & 0.091   & 0.008    \\
AE                    & 0.095   & 0.008     \\
\hline 
\end{tabular}}}
\end{center}
\caption{FC-LSTM on MovingSTL test set: prediction of 10 images when the 9 previous are given. Reconstruction MSE: MSE between raw images and reconstructed images, computed independently of the FC-LSTM predictions. Prediction MSE: MSE between raw images and reconstructed predictions. For information, the MSE obtained without using compressed representations, with a FC-LSTM trained with the same number of epochs, is 0.106.}
\label{table:lstm-STL}
\end{table}

\section{DISCUSSION}
\label{sec:discuss}
Let us recall that we wanted to compare several linear spatial latent representations for spatial compression and time prediction.

As far as spatial compression is concerned, it is not surprising that the learned representation gives the best results. Indeed, this is the only representation optimised for the task. Interestingly, in the case of MovingSTL, the semi-geometric graph is very competitive. Again, this result was expected because STL is made of natural images that are likely to be low frequency on the grid graph. In the case of MovingMNIST, the grid graph poorly captures the regularities of the graph signals, leading to a strong advantage for the learned representation. This may come from the fact that MovingMNIST images may not be smooth on the graph, as a consequence low frequencies might not capture efficiently the structure of the actual signals. Previous work on brain data has indeed shown that the high graph frequencies can also be important in capturing latent representations ~\cite{menoret2017evaluating,pilavci2019spectral}.

As far as time prediction is concerned, we observed an interest in using spatial latent representations to predict the future of sequences. Another interesting finding of our study lies in the fact that all representations perform similarly. We see three possible explanations. A first hypothesis is that FC-LSTMs trained for the task contain enough parameters to handle both spatial and temporal predictions. This would explain why the initial advantage of learned representations is lost. A second hypothesis is that by construction GFT transforms are likely to be sparse, providing a much easier task. On the contrary, we did not impose any constraints on the latent representations trained using the auto-encoder. Finally, the third hypothesis is that our FC-LSTMs are not able to capture the complex dynamics of our dataset.

Obviously, this comparison is highly theoretical. In practice, having access to the underlying spatial domain of the data, or to a larger number of samples to train the latent representation, is rare. But we found interesting that our three considered approaches ended performing so similarly in the end. The exact reasons remain unclear.

\section{CONCLUSION}
\label{sec:ccl}
In this study, we were interested in predicting spatially structured temporal sequences of data. More precisely, we wondered whether linear transforms of such data help a FC-LSTM make more accurate predictions.
We explored three linear transforms: a structured-based transform with the Graph Fourier Transform, a data-driven approach with a linear auto-encoder, and a mixed approach using a semi-geometric graph. We also introduced a dataset for this study. Our experiments confirmed the interest of a prior spatial compression of the data before predicting future frames. They also showed that domain-agnostic and data-agnostic solutions are on par on this task. In future work, we would like to investigate further the fundamental reasons why all compared methods behaved similarly. We would also like to encompass more diverse representations of data, including nonlinear variations of auto-encoders and wavelets transforms on graphs.

\appendix
\section{Predictions obtained on MovingSTL}
\label{sec:pred}
\begin{figure}[h]
  \begin{center}\includegraphics[width = 14.6cm]{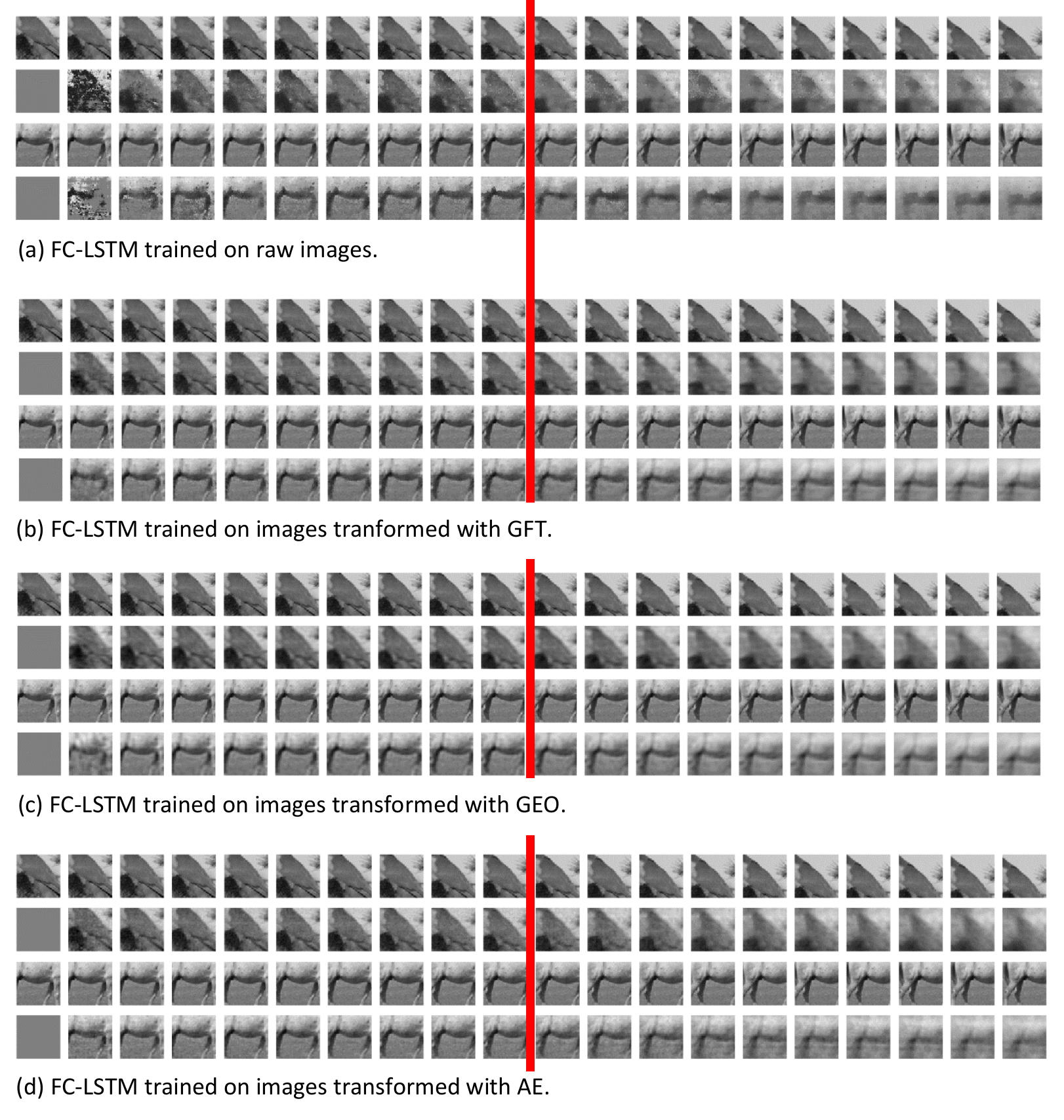}\end{center}
  \caption{\label{Fig:lstm} FC-LSTM on MovingSTL. Two examples are plotted for each method: (a) without any method, (b) with the GFT on a grid graph, (c) with the GFT on a semi-geometric graph, (d) with a linear auto-encoder. The first row of each example displays the real images. The second row displays the reconstructed output of the FC-LSTM. The 10 first images are used for warming up the FC-LSTM memory. The 10 last images (after the red line) are predictions (the FC-LSTM no longer have access to the previous real image).}
\end{figure}

\newpage
\bibliography{report} 
\bibliographystyle{spiebib} 

\end{document}